\RequirePackage{fix-cm}
\documentclass{article}
\pdfoutput=1
\usepackage{authblk}
\usepackage{graphicx}
\usepackage{tikz-qtree}
\usepackage{url}
\usepackage{array}
\usepackage{enumerate}
\usepackage{float}
\usepackage{ragged2e}
\usepackage{mathtools}
\usepackage{subcaption}
\usepackage{multirow}
\usepackage{caption}
\usepackage{longtable}
\usepackage[normalem]{ulem}
\usepackage{makecell}
\usepackage[margin=3cm]{geometry}

\newcolumntype{P}[1]{>{\centering\arraybackslash}p{#1}}
\newcommand{\bftab}{\fontseries{b}\selectfont}

\author[1]{F. Ruggeri\thanks{Corresponding Author: federico.ruggeri6@unibo.it}}
\author[2]{F. Lagioia\thanks{francesca.lagioia@eui.eu}}
\author[3]{M. Lippi\thanks{marco.lippi@unimore.it}}
\author[4]{P. Torroni\thanks{p.torroni@unibo.it}}
\affil[1, 4]{DISI, University of Bologna}
\affil[2]{CIRSFID, University of Bologna and law Department, EUI}
\affil[3]{DISMI, University of Modena and Reggio Emilia}

\begin{document}

\title{Memory networks for consumer protection: unfairness exposed
}





\maketitle

\begin{abstract}
Recent work has demonstrated how data-driven AI methods can leverage consumer protection by supporting the automated analysis of legal documents. However, a shortcoming of data-driven approaches is poor explainability. We posit that in this domain useful explanations of classifier outcomes can be provided by resorting to legal rationales. We thus consider several configurations of memory-augmented neural networks where rationales are given a special role in the modeling of context knowledge. Our results show that rationales not only contribute to improve the  classification accuracy, but  are also able to offer meaningful, natural language explanations of otherwise opaque classifier outcomes.

\end{abstract}


\section{Introduction} \label{sec:introduction}

Terms of Service (ToS), also known as terms and conditions or simply \textit{terms}, are consumer contracts governing the relation between providers and users. Terms that cause a significant imbalance in the parties’ rights and obligations, to the detriment of the consumer, are deemed \textit{unfair} by Consumer Law.
Despite substantive law in place, and despite the competence of enforcers for abstract control, providers of online services still tend to use unfair and unlawful clauses in these documents  \cite{loos2016wanted,micklitz2017empire}.
Consumers often cannot do anything about it. To begin with, they rarely read the contracts they are required to accept \cite{obar2016biggest}. Then, even if they did, a seemingly insurmountable knowledge barrier creates a clear unbalance. Legal knowledge is difficult, if not impossible, to access for individual consumers, as it is also difficult for consumers to know what data practices are implemented by companies and, therefore, to pinpoint unfair or unlawful conduct  \cite{jair2020}. Finally, even if consumers had sufficient knowledge and awareness to take legal action, there is still the insurmountable difference between the financial resources of the average customer and those of the average provider. To help mitigate such an unbalance, consumer protection organizations have the competence to initiate judicial or administrative proceedings. However, they do not have the resources to fight against each unlawful practice.
It was thus suggested that Artificial Intelligence (AI) and AI-based tools can aid consumer protection organizations and leverage consumer empowerment, for example by supporting the automatic analysis and exposure of unfair ToS clauses  \cite{lippi2019consumer}.

Among other initiatives, the CLAUDETTE project\footnote{\url{http://claudette.eui.eu/}} undertook the challenge of consumer empowerment via AI by investigating ways to automate reading and legal assessment of online consumer contracts and privacy policies with natural language processing techniques, so as to evaluate their compliance with EU consumer and data protection law.
A web service developed and maintained by the project automatically analyzes any ToS a user may feed it, and returns an annotated version of the same document, which highlights the potentially unfair clauses it contains  \cite{lippi2019claudette}.
%
While this constitutes a noteworthy first step, it suffers from poor transparency.
In other words, however accurate a system like CLAUDETTE may be, it can hardly  \textit{explain} its output. 
This shortcoming is not specific to this particular system. Indeed, in recent years a rich debate has flourished around the opacity of  AI systems that, in terms of accuracy, offer unprecedented results, but at the same time cannot be easily inspected in order to find reasons behind blatant and even possibly dangerous mistakes. This adds to the growing concern that data-driven machine-learning systems may exasperate existing biases and social inequalities  \cite{ONeil:2016:WMD:3002861,jair2020}.
The debate is very lively as it involves thinkers with all sorts of backgrounds and complementary perspectives, governments, and, to some extent, the entire civil society. 

There are good reasons for such a great interest.
Research in social science suggests that providing explanations for recommended actions deeply influences users' confidence in, and acceptance of, AI-based decisions and recommendations  \cite{cramer2008effects}. From this viewpoint, consumers, their organizations, and legal experts want to understand why a certain conclusion is made before accepting AI response. 

In our opinion, a promising approach to associating explanations to the outcomes of neural-network classifiers could be enabled by Memory-Augmented Neural Networks or MANNs  \cite{sukhbaatar2015end}. 
The basic idea behind MANNs is to combine the successful learning strategies developed in the machine learning literature for inference with a memory component that can be read and written to.
Consider for instance the following story: \begin{quote}
Joe went to the kitchen. Fred went to the kitchen. Joe picked up the milk. Joe travelled to the office. Joe left the milk. Joe went to the bathroom. \textit{Where is the milk now?} 
\end{quote}
Answering the question 
requires comprehension of the actions ``picked up'' and ``left'' 
as well as
of the time elements of the story
  \cite{weston2014memory}.
A MANN can answer these questions by storing in dedicated parts of the network, called \textit{memories},  all previously seen sentences, so as to retrieve the most relevant facts to a given query. The list of memories used to answer a given query, for example ``Joe travelled to the office'' and ``Joe left the milk'' constitutes, in a way, an explanation to the answer ``The milk is in the kitchen''.

We believe that these tasks present similarities with the problem we are tackling, of providing an explanation to why a given clause has been labeled as potentially unfair.
In particular, our hypothesis is that useful explanations may be given in terms of \textit{rationales}, i.e. ad-hoc justifications provided by legal experts motivating their conclusion to consider a given clause as unfair. Accordingly, if we train a MANN classifier to identify unfair clauses by using as facts the rationales behind unfairness labels, then a possible explanation of an unfairness prediction could be constructed based on the list of memories, i.e., the rationales, used by the MANN.
Such explanations could be especially useful to legal experts and consumers  because, rather than aiming to explain an underlying logical model or uncover the role of particular neural network connections, they would be more in line with a dialectical and communicative viewpoint, as advocated by  \cite{miller2019explanation}.

Consider for example a unilateral termination clause, giving the provider the right to suspend or terminate the service and/or the contract. In general, this provision could be unfair because, from the consumer's perspective, it could undermine the whole purpose of entering into the contract, and it may skew the balance of power between the parties involved. Indeed, the detection  of a ``\textit{unilateral termination}'' clause  ``\textit{with 98.8 percent confidence}'' could be a useful piece of information. However, the reason why a specific unilateral termination clause would be potentially unfair may not be self-evident. Instead, a more specific rationale such as ``\textit{the clause mentions the contract or access may be terminated but does not state the grounds for termination}'' could provide a more compelling argument in that regard. It would explain why a clause has been labeled as unfair, and would go in the direction of causal explanations, which are arguably more effective, in this context, than ``opaque'' confidence measures. 

This paper describes our approach to exposing unfairness by providing rationales using a MANN trained on a large corpus of online ToS.
The system we built relies on an extensive study made on  all the possible rationales associated with 5 major categories of unfair clauses, which we 
explicitly stated in the form of self-contained English sentences. This exercise served two purposes. The first one was to build a knowledge base that could help the laymen understand the possible motivations behind unfairness in the general case, and hopefully, to a broader extent, also guide service providers in defining fair terms of services. The other purpose was to be able to train MANN classifiers in detecting unfair clauses by encoding legal rationales in the memories.

The knowledge base of rationales constituted the basis for creating a corpus of 100 annotated ToS, which we used to train different MANN architecture configurations. We evaluated their performance with respect to relevant baselines, including support vector machine classifiers, convolutional neural networks and long short-memory networks. We also run an initial qualitative evaluation with domain experts in order to understand the explanatory efficacy of rationales in this context.

The novel corpus, as well as all the code needed to reproduce our experiments, are made available for research purposes\footnote{\url{https://anonymous.4open.science/r/c7070821-918c-42fb-b52e-32031d67f31a/}}.


The results on the new corpus are encouraging. The MANN architectures were able to match or outperform the baselines on all categories of unfair clauses, in some cases by a significant margin. Moreover, unlike all other baselines, the MANN could provide meaningful references to the relevant rationales, especially if during training the MANN is fed with the information of which rationales are related to which clause, a technique known as \textit{strong supervision}. These results suggest that MANN are a promising way to address the problem of explaining unfairness in consumer contracts and pave the way to their extensive use in other areas of automated legal text analytics.

The rest of this paper is organized as follows. In 
Section \ref{sec:related} we briefly discuss other machine-learning approaches in the consumer law domain, and the state of the art of machine-learning techniques used to address related problems. In 
Section \ref{sec:dataset} we describe our corpus and the rationales used to annotate it. In 
Section \ref{sec:methodology} we introduce the MANN architectures used in our study and the experimental methodology we adopted. Results are discussed in 
Section \ref{sec:results}.
Section \ref{sec:conclusion} concludes.


\section{Background} \label{sec:related}








\cite{ashley_2017} discusses a recent trend, in the legal domain, of using machine learning methods for the analysis and classification of legal documents. Recent popular solutions widely adopt traditional data-driven solutions following standard supervised learning, such as support vector machines (SVM) \cite{biagioli, lippi2019claudette, moens2007automatic}. While such approaches mainly consider detection tasks, such as argument detection \cite{lippi2015claim}, cited facts and principles in legal judgements \cite{shulayeva2017recognizing}, and prediction problems like forecasting judicial decisions \cite{aletras2016predicting}, work on consumer law is still limited, and it is mainly about consumer contracts and privacy policies \cite{fabian2017large, harkous2018polisis, sadeh2013usable, harkous2018polisis, braun2018customer,lippi2019claudette}. In particular, the last one describes a SVM classifier trained on a set of consumer contracts annotated by domain experts for clause unfairness detection. However, in all these works legal knowledge is mostly restricted to labeling information, thus, not pointing to any kind of external, accessible resource, such as legal rationales.


A topic of growing interest, not only in legal applications but in data-driven AI in general is \textit{explainability}.
Social scientists address explanation as a communication problem \cite{miller2019explanation}, concentrating on whom the explanation is provided for and thus on the interaction between explainer and explainee, as well as  on the `rules' that govern such an interaction.
Many believe that explainability (or explicability) should be an inspiring principle for the development of AI, along with beneficence, non-maleficence, autonomy and justice \cite{selbst2018intuitive, floridi2018ai4people, jobin2019artificial}. 

The communicative and dialectical dimensions of explanation are particularly relevant, even crucial, to legal experts, consumers and their organizations. In that regard, it has been argued \cite{miller2019explanation, wachter2019right} that the following approaches are needed: (i) contrastive explanation; (ii) selective explanation; (iii) causal explanation; and (iv) social explanation. While contrastive explanation is used to specify what input values determined the adoption of a certain decision (e.g., the unfairness of a clause under a certain category) rather than possible alternatives (e.g., the fairness of that specific clause), selective explanation is based on those factors that are most relevant according to human judgments. Indeed, causal chains are often too large to comprehend, especially for those who lack the specific domain competence, such as lay consumers. Causal explanation focuses on causes, rather than on merely statistical correlations. If we consider consumers, NGOs and legal experts as addressees, referring to probabilities and statistical generalizations is not as effective and meaningful as referring to causes. 
Finally, the explanation has a social nature. It is useful to adopt a conversational approach in which information is tailored to the recipient’s beliefs and way of understanding. 

Computer scientists have focused on the technical possibility of providing understandable models of opaque AI systems (and, in particular, of deep neural networks), i.e., models of the functioning of such systems that can be mastered by human experts \cite{doran2017does, guidotti2018survey}. 
%
In particular, from a computer science perspective the explanation needs to include three components. The first one is a model explanation, i.e., an \textit{interpretable} and transparent model, capturing the whole logic of the obscure system. The second is a model inspection, i.e., a human-comprehensible representation of the specific properties of an opaque system and its prediction, making it possible to understand how the black box behaves internally depending on the input values, namely its sensitivity to certain attributes (e.g., how  a change in the consumer's age, gender, location, educational level or search history  makes a difference in delivering online behavioural advertising), up to and including, for instance, the connections in a neural network. The third one is the outcome explanation, making it possible to understand the reasons for certain decisions, i.e., the causal chains leading to a certain outcome in a particular instance \cite{guidotti2018survey, arrieta2020explainable,biran2017explanation}.
%
Our approach gives special emphasis to the third component, and it consists in explaining the outcome of a classifier by integrating external knowledge expressed in natural language, in the form of rationales.


External knowledge integration in natural language processing has been widely explored in several different tasks, spanning from traditional classification \cite{wang2017combining, sun2018open} to language representation \cite{mikolov2013efficient, bian2014knowledge, zhang2019ernie, devlin2018bert} and generation \cite{zhou2018commonsense, chen2019incorporating, guan2019story}. One of the most common approaches to encode information is by means of knowledge graphs. Specifically, tasks-specific entities as well as their relations are encoded into nodes and edges in an entity-based graph, respectively. Such structured knowledge aids the detection of non-trivial connections between named entities, highlighting implicit groupings that can be used to achieve improved performance in named entity recognition \cite{callan2002knowledge, dekhili2019augmenting, torisawa2007exploiting, dadas2019combining}, sentiment analysis \cite{cambria2014senticnet, ma2018targeted, bohlouli2015knowledge, kontopoulos2013ontology, schmunk2013sentiment} and text classification \cite{wang2017combining, zelikovitz2003integrating, zhang2019integrating, choi2017gram}. Notable examples propose to enrich entity information by exploring structured ontologies, such as WordNet \cite{miller1995wordnet}, FreeBase \cite{bollacker2008freebase}, DBPedia \cite{auer2007dbpedia} and ConceptNet \cite{speer2013conceptnet}, unstructured text sources like Wikipedia as external knowledge \cite{torisawa2007exploiting, dekhili2019augmenting}, or other task specific ontologies, such as biomedical databases \cite{amith2017knowledge}.

Alternatively to knowledge graphs, prior information can be formulated as a set of constraints in the form of first-order logic (FOL) rules. Formal restrictions come as a natural representation of task specific requirements, such as structural conditions for sequential prediction tasks \cite{hu2016harnessing}. Similarly to knowledge graphs, logic rules are also employed for learning rich language representations by incorporating commonsense knowledge \cite{rocktaschel2014low, wang2014structure, bowman2014recursive, rocktaschel2015injecting}. Lastly, an interesting yet different task, termed \textit{knowledge graph embedding}, tackles the problem of encoding entire knowledge graphs into continuous vector spaces in order to convey rich external knowledge into machine learning models in an intuitive and compliant way, without losing structural properties. In this context, logic rules are employed to guide the distillation process so as to consider entity-specific relations \cite{guo2018knowledge}.



Differently from the above proposals, we relied on MANNs. 
These have been extensively employed in complex reasoning tasks, such as reading comprehension  \cite{miller2016key,hill2015goldilocks,cheng2016long}, dialogue systems \cite{DBLP:journals/corr/BordesW16} and general question answering  \cite{sukhbaatar2015end,kumar2016ask,bordes2016learning,bordes2015large}. Their novelty and main contribution is the introduction of an external memory block as support for reasoning. MANNs have also been widely applied to classification tasks, mainly sentiment analysis \cite{tang2016aspect} and document tagging \cite{DBLP:journals/corr/PrakashZHDLQLF16}, as well as graph analysis and path finding \cite{graves2016hybrid,zaremba2015reinforcement}. 

To illustrate, consider the challenging activity of question answering, where the input is usually a question concerning a single element of a corresponding text document. In this scenario, a MANN can efficiently tackle the task by encoding the question as the network input and storing the context document within the memory block. By doing so, the neural architecture can easily extract non-sequential information conditioned on the query, and potentially modify or filter the content of the memory so as to ease task resolution. Enhancing the capability of a recurrent neural network by adding the possibility of processing unordered information is an effective solution to the well-known problem of learning long-term dependencies.

By both considering unstructured knowledge and MANNs natural inclination for handling raw text, we envision arbitrary knowledge integration as an intermediate point between traditional structured information injection and natural language comprehension tasks. Thus, we find this type of memory-augmented architecture quite suitable for our purposes. A pilot study on the use of MANN for detecting and explaining unfair clauses in consumer contracts was recently presented by \cite{lagioia2019deep}, and it gave promising results. With respect to it, the present study relies on a significantly extended dataset and evaluates multiple MANN configurations (more on weak and strong supervision in Section \ref{sec:methodology}), allowing us to draw better informed conclusions.

\section{Data} \label{sec:dataset}

Our starting point is the dataset produced by \cite{lippi2019claudette},  consisting of 50 relevant online consumer contracts, i.e., Terms of Service (ToS) of online platforms, analyzed by legal experts and marked in XML. The existing annotations identify eight different categories of unfair clauses. We \textit{doubled} the size of the dataset, which now includes 100 ToS. At the same time, we narrowed down our focus to five categories. In particular, we selected the five most challenging categories and excluded from this study the remaining three categories, which were too easy, not particularly interesting, as they were almost always associated with the same rationale. 
We selected the new contracts among those offered by some of the major players in terms of global relevance, number of users, and time the service was established.\footnote{The whole corpus consists of the ToS offered by: 9gag.com, Academia.edu, Airbnb, Alibaba, Amazon, Atlas Solutions, Badoo, Betterpoints, Blablacar, Booking.com, Box, Courchsurfing, Crowdtangle, Dailymotion, Deliveroo, Deviantart, Diply, Dropbox, Duolingo, eBay, eDreams, Electronic Arts,  Endomondo, EpicGames, Evernote, Expedia, Facebook, Fitbit, Foursquare, Garmin, Goodreads, Google, Grammarly, Grinder, Groupon, Habbo, Happn, Headspace, HeySuccess, Imgur, Instagram, Lastfm, Linden Lab, LinkedIn, Masquerade, Match, Microsoft, MyHeritage, MySpace, Moves-app, Mozilla, Musically, Netflix, Nintendo, Oculus, Onavo, Opera, Paradox, Pinterest, Pokemon GO, Quora, Rayanair, Reddit,  Rovio, Shazam, Skype, Skyscanner, Slack, Snapchat, Sporcle, Spotify, Steam, Supercell, SyncMe, Tagged, Terravision, TikTok, Tinder, TripAdvisor, TrueCaller, Tumbler, Twitch, Twitter, Uber, Ubisoft, VeryChic, Viber, Vimeo, Vivino, Wechat, Weebly, WeTransfer, WhatsApp, World of Warcraft, Yahoo, Yelp, YouTube, Zalando, Zara, Zoho and Zynga.}

All the ToS we analyzed are standard terms available on the provider's website for review by potential and current consumers. Indeed, as it will be noted below, many providers assert that rather than being an obligation on the service provider to notify their users regarding changes to the ToS and even to the service, consumers are required to review the ToS from time to time, visiting the website on a regular basis to check for any changes.
The ToS collected in the dataset were gradually downloaded and analyzed over a period of eighteen months by four legal experts, with some follow-up review for the same services. Potentially unfair clauses were tagged using the guidelines described in  \cite{lippi2019claudette}.

For the purpose of this study, we have conducted an in-depth analysis of the data set and we have created a novel structured corpus of different legal rationales, with regard to the following clauses: \begin{enumerate}[$(i)$] \item liability exclusions and limitations (\texttt{ltd}); \item the provider's right to unilaterally remove consumer content from the service, including in-app purchases (\texttt{cr}) ; \item the provider's right to unilaterally terminate the contract (\texttt{ter}); \item the provider's right to unilaterally modify the contract and/or the service (\texttt{ch}); and \item arbitration on disputes arising from the contract (\texttt{a}). \end{enumerate}
Interestingly, out of the 21,063 sentences in the corpus, 2,346 sentences (11.1\%) were labeled as  containing a potentially or clearly unfair clause. We take it as a confirmation of the importance and potential impact of the analysis work we carried out. 

The distribution of the different categories across the 100 documents is reported in Table  \ref{tab:corpus_statistics}.
We shall notice the high frequency of some of the chosen categories within the dataset. 
Arbitration clauses are the most uncommon, and are found in 43 documents only. All other categories appear in at least 83 out of 100 documents. Limitation of liability and unilateral termination together represent more than half of all potentially unfair clauses. 

We expected that detecting unfair clauses under these categories would be especially challenging. Not only the state-of-the-art classifier 
showed lower performance on such clauses in comparison to other categories  \cite{lippi2019claudette},
but it also turns out that each of these potentially unfair clause categories could be matched with several potentially unfair practices/legal rationales. 
The resulting
one–to–many mapping of clauses to legal rationales will be detailed in the following sections.
The link between rationales and clauses can be used to instruct the system so that it can provide an explanation for the unfairness of particular clauses.

\begin{table}[!bt]
\begin{center}
\caption{Corpus statistics. For each category of clause unfairness, we report the overall number of clauses, the number of documents they appear in and the average sentence length (i.e. word count). \label{tab:corpus_statistics}}
\begin{tabular}{|c|c|c|c|}
    \hline
    Type of clause & \# clauses & \# documents & average length (\# words) \\
    \hline
    Limitation of liability   			     & 629  & 98 & 47.04 \\
    Content removal 		   & 216 & 83 & 39.81	\\
    Unilateral termination   		     & 420 & 93 & 36.04	\\
    Unilateral changes 			     & 344  & 97 & 26.03	\\
    Arbitration  & 106 & 43 & 47.55 \\
    \hline
  \end{tabular}
\end{center}
\end{table}

\subsection{Limitation of liability clauses}
 
Service providers often dedicate a considerable portion of their ToS to disclaiming and limiting liabilities.  
Clauses falling under this category stipulate that the duty to pay damages is limited or excluded for certain kinds of losses and under certain conditions. Most of the circumstances under which these limitations are declared significantly affect the balance between the parties' rights and obligations, and unlikely will pass the Directive's unfairness test. 
In particular, clauses excluding liability for broad categories of losses or causes of them were  marked as potentially unfair, including those containing blanket phrases like ``to the fullest extent permissible by law''. Conversely, clauses meant to reduce, limit, or exclude the liability for physical injuries, intentional harm, or gross negligence were marked as clearly unfair  \cite{lippi2019claudette,micklitz2017empire}.

The analysis of the dataset enabled us to identify 19 legal rationales for (potentially) unfair limitation of liability, which map different questionable circumstances under which the ToS reduce or exclude liability for losses or injuries. 
For each rationale we defined a corresponding identifier [ID], as shown in Table  \ref{tab:LTDlegal_rationales}.

    \begin{scriptsize}
    \begin{longtable}{|p{2.5cm}|p{8.5cm}|} 
    \caption{Legal rationales for the legal qualification of limitation of liability unfairness.\label{tab:LTDlegal_rationales}}
    \\ \hline
        \textbf{ID} & \textbf{Legal Rationale} \\ \hline
        \texttt{extent}  & \texttt{Except as required by law, or to the fullest extent permissible by applicable law the provider is not liable, and/or  users are solely responsible for ensuring that the Terms of Use/Service are in compliance with all laws, rules and regulations, and the use of the platform its on their own risk.} \\ \hline 
        \texttt{gen} & \texttt{The clause introduces a general and non-specific  limitation and/or exclusion of liability,   such as  liability for various things, liability arising out of or in connection with the service and/or the Terms.}\\ \hline
        \texttt{discontinuance}  & \texttt{The provider is not liable for any technical problems, failure, inability to use the service, suspension, disruption, modification, discontinuance, unavailability of service, any unilateral change, unilateral termination,  unilateral limitation  including  limits on certain features and services or restriction to  access to parts or all of the Service without notice } \\ \hline 
        \texttt{compharm} & \texttt{The provider is not liable for  harm or damage to hardware and software, including viruses, malware, worms, trojan horses, or any similar contamination or destructive program.} \\ \hline 
        \texttt{anydamage}  & \texttt{The provider is not liable for any special, direct and/or indirect, punitive, incidental or consequential  damage, including negligence, and broad categories of damages, including harm or failure.} \\ \hline
       \texttt{amount} & \texttt{The compensation for liability or aggregate liability is limited to, or should not exceed, a certain total amount, or that the sole remedy is to stop using the service and cancel the account, or that you can't recover any  damages or losses.} \\ \hline 
       \texttt{thirdparty}  & \texttt{The provider is not liable for any action, errors, omissions, representations, warranties, breaches or negligence  taken from third parties, third-party providers services, suppliers or other people, acts of any government and authority, including service and products, additional costs, copyright compliance, legality or decency of material, content and link  posted by others, including users.} \\ \hline 
       \texttt{security}  & \texttt{The provider is not liable for any damage deriving from a security breach, including any unauthorised access, alteration and modification of  data, data transmission.} \\ \hline  
       \texttt{dataloss}  & \texttt{The provider is not liable for any disclosure, damage, destruction, corruption, failure to store or loss of data and material.} \\ \hline 
       \texttt{reputation}  & \texttt{The provider is not liable for reputational  and goodwill  damages, loss.} \\ \hline 
       \texttt{anyloss}  & \texttt{The provider is not liable for any loss , or broad categories of loss, resulting from the use of the service and or of the website, including lost profits, lost opportunity, lost business or lost sales, data loss, loss of goodwill, loss of reputation.} \\ \hline 
       \texttt{awareness} & \texttt{The provider is not liable even if he was, or should have been, aware or have been advised about the possibility of any damage or loss.} \\ \hline 
       \texttt{contractfailure}  & \texttt{The provider is not liable for any failure in performing contract and terms, obligations, including unavailability or failure in providing products and services, breach of agreement, lack of performance.} \\ \hline 
      \texttt{dataloss} & \texttt{The provider is not liable for any loss of data.} \\ \hline 
      \texttt{ecoloss} & \texttt{The provider is not liable for any loss of  profits, loss of income, lost opportunity, lost business or lost sales, loss of revenue.} \\ \hline
      \texttt{content}  & \texttt{The provider is not liable for any information stored or processed within the Services, inaccuracies or error of information, content and material posted, software, products and services on the website, including copyright violation, defamation, slander, libel, falsehoods, obscenity, pornography, profanity, or objectionable material.} \\ \hline 
      \texttt{liabtheories}  & \texttt{The provider is not liable under different theories of liability, including tort law, contract law,  strict liability, statutory liability, product liability and other liability theories.} \\ \hline 
       \texttt{grossnegligence}  & \texttt{The provider is not liable for gross negligence.} \\ \hline 
        \texttt{injury}  & \texttt{The provider is not liable for intentional offence and damage, physical or personal injury, including, emotional distress and  death.} \\ \hline
    \end{longtable}
   \end{scriptsize} 

As noted above, each (potentially) unfair \texttt{ltd} clause within the data set may be relevant, and thus indexed with the corresponding IDs, 
under more than one legal rationale. 
As an example, consider the following clause taken from the Box terms of service (retrieved on August 2017) 
and classified as potentially unfair:
\begin{quotation}

\begin{sloppypar}
\texttt{To the extent not prohibited by law, in no event will Box, its affiliates resellers, officers, employees, agents, suppliers or licensors be liable for: any direct, incidental, special, punitive, cover or consequential damages (including, without limitation, damages for lost profits, revenue, goodwill, use or content) however caused, under any theory of liability, including, without limitation, contract, tort, warranty, negligence or otherwise, even if Box has been advised as to the possibility of such damages.}
\end{sloppypar}

\end{quotation}

The clause above, to the extent not prohibited by law, limits the provider's liability by kind of damages, i.e., broad category of losses (e.g. loss of data, economic loss and loss of reputation); by standard of care, since it states that the provider will never be liable even in case of negligence and awareness of the possibility of damages; by causal link (e.g. special, incidental and consequential damages) as well as by any liability theory. 
As a consequence, the clause has been linked to the following identifiers: \texttt{extent, anydamage, reputation, dataloss, ecoloss, awareness, liabtheories}.

As a further example, consider the following clause taken from the Endomondo terms of service (retrieved on May 2018) and classified as clearly unfair:

\begin{quotation}

\begin{sloppypar}
\texttt{Except as otherwise set out in these Terms, and to the maximum extent permitted by applicable law, we are not responsible or liable, either directly or indirectly, for any injuries or damages that are sustained from your physical activities or your use of, or inability to use, any Services or features of the Services, including any Content or activities that you access or learn about through our Services (e.g., a Third-Party Activity such as a yoga class), even if caused in whole or part by the action, inaction or negligence of Under Armour or by the action, inaction or negligence of others.}
\end{sloppypar}
\end{quotation}

To the maximum extent permitted by the law, the clause above limits the service provider's liability by the causal link with broad categories of potential damages (i.e., to encompass direct and/or indirect damages), by cause (i.e., service interruption and/or unavailability, third party actions and content published,  stored, and processed within the service), and by kind, in particular for personal injury, resulting from an act or an omission of the supplier. Thus, it has been linked to the following identifiers: \texttt{extent, anydamage, injury, thirdparty, content, discontinuance}.

\subsection{Content removal}

Content removal clauses give the provider a right to modify and/or delete user's content, including in-app purchases, and sometimes specifies the conditions under which the service provider may do so.
As in the case of unilateral termination, clauses that indicate conditions for content removal were marked as potentially unfair, whereas clauses stipulating that the service provider may remove content in his full discretion, and/or at any time for any or no reasons and/or without notice nor possibility to retrieve the content were marked as clearly unfair.

Under this category, we identified 17 legal rationales, shown in Table  \ref{tab:CRlegal_rationales}.

\begin{scriptsize}
    \begin{longtable}{|p{2.5cm}|p{8.5cm}|}
    \caption{Legal rationales for the Content Removal category.\label{tab:CRlegal_rationales}}
    \\ \hline
        \textbf{ID} & \textbf{Legal Rationale} \\ \hline
        \texttt{nonotice}  & \texttt{The provider has the right to remove content and material without prior notice} \\ \hline 
        \texttt{noretrieve}  & \texttt{The provider has the right to remove content and material without the possibility to retrieve such content and material.} \\ \hline 
        \texttt{fulldiscretion} & \texttt{The provider has the right to remove content and material for any reason, at its full discretion.} \\ \hline 
        \texttt{lawviolation}  & \texttt{The provider has the right to remove content and material in order to comply with applicable law, if he believes  in good faith that there is a case of law violation, including intellectual property infringments.} \\ \hline
       \texttt{termviolation} & \texttt{The provider has the right to remove content and material if he believes that there is a case  violation of terms such as acount tranfer, policies, standard, code of conduct.} \\ \hline       
       \texttt{objcontent}  & \texttt{The provider has the right to remove content and material that he considers to be offensive, obscene, abusive, harmful, objectable, inaccurate, inappropriate.} \\ \hline 
       \texttt{comply}  & \texttt{The provider has the right to remove content and material in order to comply with  the order or request of a court, law enforcement , other administrative agency or governmental body.} \\ \hline 
       \texttt{serviceprotection}  & \texttt{The provider has the right to remove content and material that he considers to be harmful to  or as creating threats for the provider's property, Site or Services, or consumers.} \\ \hline 
       \texttt{criteriafailure}  & \texttt{The provider has the right to remove content and material if there is a failure in meeting  any applicable quality or eligibility criteria.} \\ \hline 
       \texttt{complaint}  & \texttt{The provider has the right to remove content and material in case of complaints about users' performance, conduct, published content and information.} \\ \hline 
       \texttt{rating}  & \texttt{The provider has the right to remove content and material in case of poor Ratings or Reviews.} \\ \hline 
       \texttt{fraudprevention}  & \texttt{The provider has the right to remove content and material in order to prevent fraud and illegal activities.} \\ \hline 
       \texttt{personalsafety}  & \texttt{The provider has the right to remove content and material to protect personal safety.} \\ \hline
       \texttt{liability}  & \texttt{The provider has the right to remove content and material if they could subject the provider to liability.} \\ \hline
       \texttt{tpright}  & \texttt{The provider has the right to remove content and material if they constitute a violation of third party rights, including trademarks.} \\ \hline
       \texttt{susp}  & \texttt{The provider has the right to remove content and material upon suspension or termination.} \\ \hline
       \texttt{inactive}  & \texttt{The provider has the right to remove content and material in case of users' inactivity.} \\ \hline
    \end{longtable}
   \end{scriptsize} 

Each (potentially) unfair clause falling under the content removal category has been indexed with one or more identifiers of rationales.
Consider, for instance, the following examples 
 taken from the terms of service of TikTok (retrieved on October 2018) and 
 Pokemon GO (retrieved on July 2016): 

\begin{quotation}
\begin{sloppypar}
\texttt{In addition, we have the right - but not the obligation - in our sole discretion to remove, disallow, block or delete any User Content (i) that we consider to violate these Terms, or (ii) in response to complaints from other users or third parties, with or without notice and without any li\-a\-bil\-i\-ty to you.}
\end{sloppypar}
\end{quotation} 
 
\begin{quotation}
\begin{sloppypar}
\texttt{Niantic further reserves the right to remove any User Con\-tent from the Services at any time and without notice and for any reason.}
\end{sloppypar}
\end{quotation}

The first clause above, previously classified as potentially unfair,  has been linked to the following IDs: \texttt{termviolation, complaint, nonotice, tpright}.
Coversely, the second clause, previously classified as clearly unfair,  has been linked to the IDs \texttt{IDs: nonotice, fulldiscretion}.

\subsection{Unilateral termination}

The unilateral termination clause gives the provider the right to suspend and/or terminate the service and/or the contract, and sometimes details the circumstances under which the provider claims to have a right to do so. 
From the consumer's perspective, a situation where the agreement may be dissolved at any time and for any reason could seriously undermine the whole purpose of entering into the contract.
These clauses may skew the balance of power between the parties and be considered (potentially) unfair whenever the consumer has a reasonable interest in preserving the contract's longevity, given the foreseeably invested time and effort in the services, e.g., by importing and storing digital content. 
This is all the more true if the trader does not provide a reasonably long notice period allowing the consumer to migrate to another service (e.g., withdrawing and transferring all the digital content elsewhere).

Unilateral termination clauses that specify reasons for termination were marked as potentially unfair, whereas clauses stipulating that the service provider may suspend or terminate the service at any time for any or no reasons and/or without notice were marked as clearly unfair. 
Under this (potentially) unfair clause category, we identified 28 different legal rationales.

\begin{scriptsize}
    \begin{longtable}{|p{2.5cm}|p{8.5cm}|}
    \caption{Legal rationales for the legal qualification of Unilateral Termination unfairness.\label{tab:TERlegal_rationales}}
    \\ \hline
        \textbf{ID} & \textbf{Legal Rationale} \\ \hline
        \texttt{fraud\_abuse\_illegal}  & \texttt{The contract or access may be terminated where the user has been engaging in illegal or unlawful activity, including fraudulent behaviour, abusive, misusive or otherwise harmful behaviour, or for reasons of safety or fraud prevention.} \\ \hline 
        \texttt{breach}  & \texttt{The contract or access can be terminated where the user fails to adhere to its terms, or community standards, or the spirit of the ToS or community terms, including inappropriate behaviour, using cheats or other disallowed practices to improve their situation in the service, deriving disallowed profits from the service, or interfering with other users' enjoyment of the service or otherwise puts them at risk, or is investigated under any suspision of misconduct.} \\ \hline 
        \texttt{no\_grounds} & \texttt{The clause mentions the contract or access may be terminated but does not state the grounds for termination.} \\ \hline 
        \texttt{misinfo}  & \texttt{The clause mentions the contract or access may be terminated where the user has provided false, outdated or incomplete information.} \\ \hline
       \texttt{infring\_tp\_rights} & \texttt{The contract or access may be terminated in cases of  infringement upon rights of others, including copyrights or other intellectual property rights, including termination for repeat infringers.} \\ \hline 
       \texttt{multiple}  & \texttt{The contract or access may be terminated in cases of a single user holding or controlling multiple accounts, or multiple use of a single account.} \\ \hline 
       \texttt{cred\_security}  & \texttt{The contract or access may be terminated where the user fails to maintain the security of the login credentials and/or a security breach occurs.} \\ \hline 
       \texttt{dormant}  & \texttt{The contract or access may be terminated where the account has been left dormant for a prescribed time.} \\ \hline
       \texttt{user\_bad\_rep}  & \texttt{The contract or access may be terminated where the user fails to maintain a prescribed level of reputation.} \\ \hline
       \texttt{reference}  & \texttt{The contract or access may be terminated but refers to grounds formulated elsewhere.} \\ \hline
       \texttt{content\_violation}  & \texttt{The contract or access may be terminated where the user has entered content into the service which is, or is deemed to be, infringing upon the rights of others or in violation with the terms of service.} \\ \hline
        \texttt{payments}  & \texttt{The contract or access may be terminated where the user has not been meeting their payment obligations, or withdrawing payments, e.g. via chargeback.} \\ \hline
        \texttt{gen\_rights\_violation}  & \texttt{The clause generally states the contract or access may be terminated where the user has violated the rights of the service provider or other entities.} \\ \hline
         \texttt{over\_quota}  & \texttt{The contract or access may be terminated where the user has been violating the time, storage or other limits of the service.} \\ \hline
         \texttt{insolvency}  & \texttt{The contract or access may be terminated where one of the parties has been declared insolvent, bankrupt, has a court receiver or a similar officer appointed, or proceedings are pending in regard to any of these.} \\ \hline
         \texttt{shutdown}  & \texttt{The contract or access may be terminated where the service is being shutdown or ceases to be available for any other reasons.} \\ \hline
         \texttt{no\_consent}  & \texttt{The contract or access may be terminated where the user's consent is missing or withdrawn, or where the user otherwise objects to the terms, policy or any change thereof.} \\ \hline
         \texttt{sex\_of}  & \texttt{The contract or access may be terminated where the user is a registered sex offender, or engaged or attempted to engage in sexual conduct with minors, or has been involved with child pornography.} \\ \hline
         \texttt{parole}  & \texttt{The contract or access may be terminated where the user has engaged in a parole or probation violation.} \\ \hline
         \texttt{viability\_eligibility}  & \texttt{The contract or access may be terminated where the provision of the service to the user is no longer economically viable, or where the user is not eligible for the service.} \\ \hline
         \texttt{dispute}  & \texttt{The contract or access may be terminated where the user engages in a dispute with the service provider or owner.} \\ \hline
         \texttt{legal\_reasons}  & \texttt{The contract or access may be terminated to comply with legal requirements, or as a result of a request put in by authorities, or for broadly specified legal reasons.} \\ \hline
         \texttt{tech\_reasons}  & \texttt{The clause broadly states the contract or access may be terminated for technical reasons.} \\ \hline
         \texttt{any\_reasons}  & \texttt{The clause generally states the contract or access may be terminated for any reason, without cause or leaves room for other reasons which are not specified.} \\ \hline
         \texttt{force\_majeure}  & \texttt{The clause generally states the contract or access may be terminated in an event of a force majeure, act of God or other unforeseen events of a similar nature.} \\ \hline
         \texttt{providers\_exposure}  & \texttt{The contract or access may be terminated where the user's actions or content create a risk of legal exposure, or damage to the provider's reputation.} \\ \hline
         \texttt{protect\_rights}  & \texttt{The clause generally states the contract or access may be terminated to protect the rights and/or interests of the service provider or a third party.} \\ \hline
         \texttt{no\_notice} & \texttt{The clause generally states that the contract or access may be terminated without notice or simply posting it on the website and/or the trader is not required to observe a reasonable period for termination.} \\ \hline
    \end{longtable}
   \end{scriptsize} 
   
As examples, consider the  following clauses, taken from the DeviantArt (effective date not available) and Academia (retrieved on May 2017) terms of service: 

\begin{quotation}
\begin{sloppypar}
\texttt{Furthermore, you acknowledge that DeviantArt reserves the right to terminate or suspend accounts that are inactive, in DeviantArt's sole discretion, for an extended period of ti\-me (thus deleting or suspending access to your Content)}
\end{sloppypar}
\end{quotation}

\begin{quotation}
\begin{sloppypar}
\texttt{Academia.edu reserves the right, at its sole discretion, to discontinue or terminate the Site and Services and to ter\-mi\-na\-te these Terms, at any time and without prior notice.}
\end{sloppypar}
\end{quotation}

The first clause above, previously classified as potentially unfair, has been linked to the 
\texttt{dormant}
ID, since it states that the service provider will suspend or terminate apparently dormant accounts, also deleting or suspending access to consumers' digital contents.
Conversely, the second clause was previously classified as clearly unfair and it has been linked to the IDs 
\texttt{no\_grounds, no\_notice}.
since the service provider claims the right to unilaterally terminate both the contract and the service without prior notice and grounds for termination are completely missing.

\subsection{Unilateral changes}

Under this category, we identified 7 different legal rationales, as reported in Table  \ref{tab:CHlegal_rationales},  mapping the different circumstances under which service providers claim their right to unilaterally amend and modify the terms of service and/or the service. Unilateral change clauses were always considered as potentially unfair, since the ECJ has not yet issued a judgment in this regard, though the Annex to the Directive contains several examples supporting such a qualification.
Unilateral change clauses are particularly worrisome in cases where the proposed amendment significantly impact the consumers' rights,  thus creating a disproportionate balance between the parties. This is particularly true whenever consumers are either not given any opt-out options, where no consent to the new terms is requested, or where no notification to the consumers is given. 
 
\begin{scriptsize}
    \begin{longtable}{|p{2.5cm}|p{8.5cm}|}
    \caption{Legal rationales for the Unilateral Change category.\label{tab:CHlegal_rationales}}
    \\ \hline
        \textbf{ID} & \textbf{Legal Rationale} \\ \hline
        \texttt{anyreason}  & \texttt{The provider has the right for unilateral change of the contract/services/goods/features for any reason at its full discretion, at any time.} \\ \hline 
        \texttt{nowarning}  & \texttt{The provider has the right for unilateral change of the contract/services/goods/features with no notice to the consumer .} \\ \hline 
        \texttt{justposted} & \texttt{The provider has the right for unilateral change of the contract/services/goods/features where the notification of changes is left at a full discretion of the provider such as by simply posting the new terms on their website without a notification to the consumer.} \\ \hline 
       \texttt{consresp} & \texttt{The provider has the right for unilateral change of the contract/services/goods/features  where it is the consumer's responsibility to regularly check the terms for any updates.} \\ \hline 
       \texttt{againsterms}  & \texttt{The provider has the right for unilateral change of the contract/services/goods/features if the consumer violates the Terms (as a consequence only limited or no services might be provided).} \\ \hline 
       \texttt{lawchange}  & \texttt{The provider has the right for unilateral change of the contract/services/goods/features to reflect  changes in law, regulatory requirements at their own discretion.} \\ \hline 
       \texttt{update}  & \texttt{The provider has the right for unilateral change of the contract/services/goods/features to maintain a level of flexibility to amend and update services, including discontinuation.} \\ \hline 
    \end{longtable}
   \end{scriptsize} 
   
As relevant examples, consider the following clauses taken from the Acad\-e\-mia (retrieved on May 2017) and Endomondo (retrieved on January 2016) terms of service:
  
  \begin{quotation}
\begin{sloppypar}
\texttt{Academia.edu reserves the right, at its sole discretion, to modify the Site, Services and these Terms, at any time and without prior notice.}
\end{sloppypar}
\end{quotation}

\begin{quotation}
\begin{sloppypar}
\texttt{With new products, services, and features launching all the time, we need the flexibility to make changes, impose limits, and occasionally suspend or terminate certain offerings.}
\end{sloppypar}
\end{quotation}

The first clause above is representative of the largest group of the unilateral change clauses, stating that the provider has the right for unilateral change of the
contract and/or the services and/or the provided goods and/or features, for any reason, at its full discretion and at any time and without notice. Thus, it has been linked to the following IDs: \texttt{anyreason, nowarning}.
The second clause, stating that the service provider has the right to make generic unilateral changes to maintain a level of flexibility, has been linked to the ID \texttt{update}.

\subsection{Arbitration}

The arbitration clause can be considered as a kind of forum selection clause, since it requires or allows the parties to resolve their disputes through an arbitration process, before the case could go to court. 
However, such a clause may or may not specify that arbitration should occur within a specific jurisdiction. While clauses defining arbitration as fully optional has been marked as clearly fair, those stipulating that the arbitration should (1) take place in a state other than the state of consumer's residence and/or (2) be based not on law but on arbiter's discretion were marked as clearly unfair. In every other case, the arbitration clause was considered as potentially unfair. 

Under this category, we identified 8 legal rationales, as reported in Table  \ref{tab:Alegal_rationales}.

\begin{scriptsize}
    \begin{longtable}{|p{2.5cm}|p{8.5cm}|} 
    \caption{Legal rationales for the Arbitration  category.\label{tab:Alegal_rationales}}
    \\ \hline
        \textbf{ID} & \textbf{Legal Rationale} \\ \hline
        \texttt{arb\_obligatory}  & \texttt{All disputes must be resolved through arbitration, instead of a court of law, and the rights and obligations of the party will be decided by an arbitrator instead of a judge or jury.} \\ \hline 
        \texttt{exceptions\_apply}  & \texttt{Arbitration is mandatory though the clause contains exceptions where arbitration is not mandatory or does not apply under certain circumstances; this includes pursuing certain claims in a small claims court.} \\ \hline 
        \texttt{extralegal\_rules} & \texttt{The consumer is mandatorily subject to rules on dispute resolution not covered by law; this includes any rules on arbitration coined by an arbitral body, chamber, association or other type of organization.} \\ \hline 
        \texttt{outside\_domicile}  & \texttt{The arbitration is to take place in country different than the consumer's domicile.} \\ \hline
       \texttt{opt\_out} & \texttt{The consumer must first opt out for the arbitration not to be obligatory.} \\ \hline 
       \texttt{unless\_prohibited}  & \texttt{The arbitration is mandatory unless prohibited by applicable law.} \\ \hline 
       \texttt{soft\_redirect}  & \texttt{Disputes which are unresolved informally, through a small claims court or otherwise, may be submitted for arbitration, also on option of one of the parties.} \\ \hline 
       \texttt{consent\_tos}  & \texttt{The user is bound by the arbitration clause on grounds of accepting the ToS.} \\ \hline 
    \end{longtable}
   \end{scriptsize} 


As examples, consider the following clauses taken from the Airbnb (retrieved on November 2019) and Grindr (retrieved on July 2018) terms of service: 

\begin{quotation}
\begin{sloppypar}
\texttt{By accepting these Terms of Service, you agree to be bound by this arbitration clause and class action waiver.}
\end{sloppypar}
\end{quotation}

\begin{quotation}
\begin{sloppypar}
\texttt{Any Covered Dispute Matter must be asserted individually in binding arbitration administered by the American Arbitration Association (AAA) in accordance with its Consumer Arbitration Rules (including utilizing desk, phone or video conference proceedings where appropriate and permitted to mitigate costs of travel).}
\end{sloppypar}
\end{quotation}

The first clause above, previously classified as potentially unfair, has been linked to the 
\texttt{consent\_tos} 
identifier, since it states that the agreement to the Terms of Service is specifically treated as consent to the arbitration clause.
Conversely, the second clause was previously classified as clearly unfair and it has been linked to the following IDs: 
\texttt{arb\_obligatory, extralegal\_rules} 
since it states that  all disputes must be resolved through arbitration, and the consumer is mandatorily subject to rules on dispute resolution not covered by law, i.e. Consumer Arbitration Rules by the American Arbitration Association (AAA).


\section{Methodology} \label{sec:methodology}



From a machine learning perspective, it is usually hard to endow classification models with the ability to produce interpretable results, resembling justifications such as those provided by experts. With deep neural models, interpretability becomes even harder. Taking inspiration from experts' behaviour, we aim to enrich general text classification by emulating the aforementioned perspective on rationales. Specifically, implicit evaluation is addressed by data-driven learning, whereas explicit comparison with available external knowledge is modelled at architectural level. In particular, we view legal rationales as our knowledge base (KB) and we focus on their linking to given statements that have to be classified, as a way to explicitly address model interpretability.

Differently from common knowledge integration techniques in natural language processing \cite{hu2016harnessing, dhingra2016gated}, justifications given by experts cannot be easily formalized as structured information, such as knowledge graphs or logic rules, unless a high pre-processing effort is carried out. This is mainly due to the presence of abstract concepts, implicit references to several external sources, and common sense motivations. Additionally, such aspects may also be deeply intertwined and, thus, hard to properly isolate. Nonetheless, word-level information, e.g. available in ontologies \cite{miller1995wordnet}, and context-grounded statements, such as unfair examples listing, can be easily formalized and integrated into the learning phase. Intuitively, the former unstructured knowledge representation, in the form of raw text, is a generalization of the latter structured formalization, achievable via ad-hoc distillation techniques, either manually or, more ambitiously, automatically. Therefore, we focus on integrating external information, i.e., legal rationales, in order to both convey rich contextual reasoning as well as providing a simple interface to account for model interpretability. 

\subsection{Memory-augmented neural networks}

As a first building block, our choice fell on MANNs due to their extensive usage in tasks that present similarities with our approach, such as machine reading comprehension and question answering. 
We could choose among several MANN architectures of various levels of sophistication. However, since our main motivation concerns correct knowledge usage without negatively affecting the overall classification performance, it turned out that the simplest MANN architecture available \cite{sukhbaatar2015end} did met our needs. We may explore alternative, more advanced and typically much more computationally expensive architectures in future developments.

We formulate the task of potentially unfair clause detection as a standard classification problem, yet enhanced by introducing an external knowledge, containing possible explanations for the labeled unfairness types.
Specifically, we extend the setup described by \cite{lagioia2019deep}, in which statements to be classified are explicitly compared with the given knowledge. In particular, following the technical formulation of MANNs, we consider input clauses as a query to the memory, which contains a collection of legal rationales.

\begin{figure}
  \begin{subfigure}{\linewidth}
    
    \includegraphics[width=\linewidth]{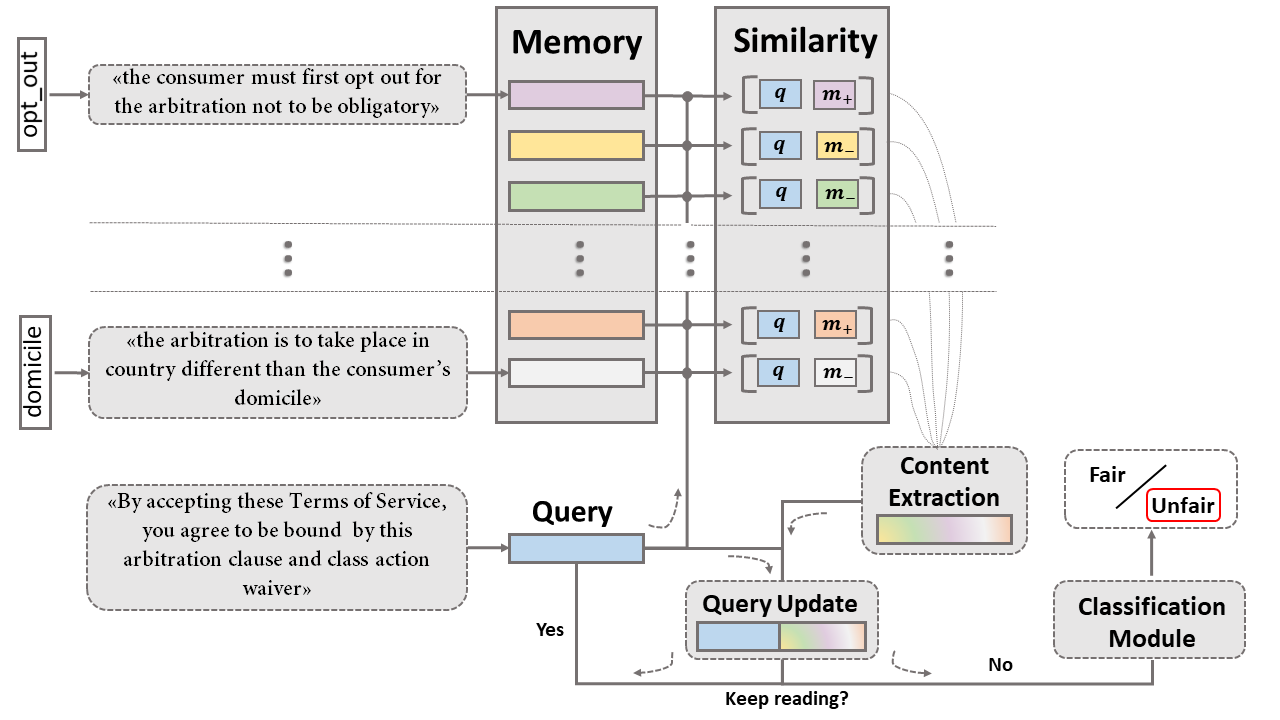}
    \caption{} 
    \label{fig:model_architecture_A}
  \end{subfigure}%
  
  \hspace*{\fill}   
  
  \begin{subfigure}{1.0\textwidth}
    \centering
    \includegraphics[width=0.67\linewidth]{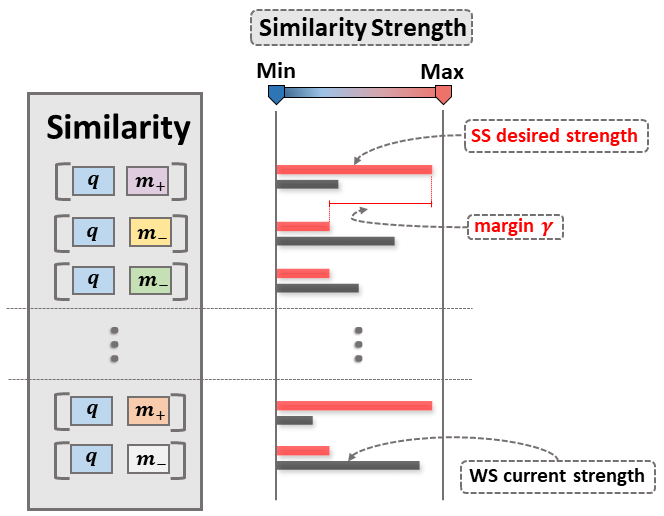}
    \caption{}
    \label{fig:model_architecture_B}
  \end{subfigure}%
  
  \caption{(a) MANN general architecture for unfair clause detection. Given a clause $q$ to classify and a set of legal rationales, i.e., memory $M$, the model first evaluates the similarity between $q$ and each memory slot $m_i$. Subsequently, proportionally to their similarity with $q$, memory components $m_i$ are aggregated into a single vector representation. Intuitively, if the similarity metric is non-zero only for a single $m_i$, the aggregated vector is a weighted representation of $m_i$. Lastly, such aggregated vector is used to update $q$ so as to enrich input information for the final classification step.\\ (b) Weak and strong supervision. The parametric similarity step can be guided by specifying which memory content $m_i$ should have high similarity with $q$. This process is formally known as strong supervision (SS) in the MANN literature. In particular, the model is instructed to shift from its current similarity behaviour (black bars) to the one desired by SS (red bars).}
\end{figure}

When a clause $q$ has to be classified, the system attempts to retrieve from memory $M$ those rationales (possibly more than one) that best match such clause. This match can be computed by exploiting a similarity metric $s(q,m_i)$ between the input clause and each memory slot $m_i$. Then, the MANN extracts content from memory $M$, by computing a weighted combination of all the memory slots, where the weight $w_i$ of the $i$-th slot is proportional to the similarity metric: $c = \sum_{i=1}^{|M|} w_i \cdot m_i$. In our setting, we compute weight $w_i$ by applying a sigmoid function $\sigma$ to the output of the similarity function: $w_i = \sigma(s(q,m_i))$.
%
Content $c$ is then combined with clause $q$ to build the final representation $\tilde{q}$ to be used to classify the clause. In our experimental setting this query update consists in a simple concatenations of the two vectors.\footnote{The general MANN architecture would also allow for multiple iterations of memory reading operations. We leave to future research a deep investigation of this possibility.}
The MANN finally considers $\tilde{q}$ to predict whether the input clause is potentially unfair.
The whole architecture of the system is depicted in Figure \ref{fig:model_architecture_A}.

\subsection{Weak and strong supervision}

We consider two different kinds of supervisions for the use of memory, usually named \textit{weak} and \textit{strong} supervision \cite{sukhbaatar2015end}.
Under weak supervision we feed the MANN with the whole collection of rationales (the KB), without providing the information, during training, of which set of rationales should be used for which clause. Under strong supervision, instead, we provide the MANN with the explanation(s) used for every potentially unfair clause.

Strong supervision of legal rationales encoded in the memory is technically implemented as a max margin loss at extraction level, and it can be informally envisioned as suggesting higher preference for memory elements, i.e., legal rationales, that are labeled by experts as being the true motivation of given clause unfairness.
This added penalty term, to be minimized, gives a high cost to examples (unfair clauses) that are not assigned to the corresponding true explanation(s) in the memory:\footnote{It is worthwhile noticing that  each clause in the training set could be associated with multiple justifications. We consider the memory to be correctly used if at least one of the supervision rationales is selected.}

\begin{equation}
    \mathcal{L}_{SS} = \frac{1}{N} \sum_{n=1}^{N} \frac{1}{|M^n_+| |M^n_-|} \sum_{m_+ \in M^n_+} \sum_{m_- \in M^n_-} \left [ \mathcal{L} (m_+, m_-) \right ] \\
\end{equation}
\begin{equation}
    \mathcal{L} (m_+, m_-) =  \max \Big (0, \gamma - \sigma(s(q^n, m_+)) + \sigma(s(q^n, m_-)) \Big )
\end{equation}
where $M^n_+$ is the set of target explanations for a given example $q^n$, $M^n_-$ is the set of non-target explanations for $n$, $(m_+, m_-)$ is a target/non-target explanation pair for a given sample $n$, and $\gamma$ is the margin hyper-parameter. Intuitively, this loss function pushes the MANN to compute scores for the target explanations that are larger than those for non-target explanations for at least a margin $\gamma$.
By controlling the intensity of this preference via a sort of regularization coefficient, we can trade-off between classification performance and model output interpretability (see Figure \ref{fig:model_architecture_B}).
In fact, in our scenario, being able to properly motivate the proposed model predictions is as crucial as evaluating its performance by standard classification metrics. As already mentioned, the introduction of a KB comprised of expertise rationales allows us to better interpret the model output by evaluating its linkage to the KB.

\section{Results} \label{sec:results}




For each unfairness category introduced in Section \ref{sec:dataset} we employed sets of expert justifications, each of a different size. In particular, we consider each category as a standalone binary classification task, where model evaluation is defined as a standard 10-fold cross-validation.
As it is a well known fact that standard MANN architectures are affected by high variance \cite{weston2014memory}, for each fold in the cross-validation three networks are trained with different random seed initializations, picking the best according to the performance on the validation set, as customary in many applications \cite{bengio2012practical}.

All models are implemented in Tensorflow 2 and are available for reproducibility\footnote{\url{https://anonymous.4open.science/r/c7070821-918c-42fb-b52e-32031d67f31a/}}.

As for the deep architecture, we employ a simple variant of the traditional end-to-end MANN \cite{sukhbaatar2015end}, with the following modifications, already introduced in Section \ref{sec:methodology}:
\begin{enumerate}
    \item attention operation via sigmoid activation function instead of softmax in order to enable the selection of multiple (possibly none) memory slots;
    \item number of reading iterations over memory limited to just one, which also simplifies training;
    \item query update via concatenation instead of embedding sum;
    \item clause-justification similarity operation via a two-layer MLP, to further take into account strong supervision as an additional penalty term. 
\end{enumerate}
The last two architectural choices follow a preliminary experimental evaluation and make the model more expressive.

For each test fold in the cross-validation, the remaining data are split between 80\% for training and 20\% for validation. Training is then regularized by early stopping on validation $F_1$, with a patience equal to 40 epochs.

As for performance comparison, we consider the current state-of-the-art solution implemented in the CLAUDETTE system \cite{lippi2019claudette}, based on support vector machines (SVM), as well as a set of deep neural networks that do not leverage any kind of external information: (i) a simple set of stacked convolutional neural networks (CNNs), on the top of which a 2-layer MLP classifier is added for binary classification; (ii) a set of stacked recurrent neural networks along with a 2-layer MLP for classification.

We set the hyper-parameters of each model, including MANNs, by selecting the architecture with the best performance on the validation set. In particular, for MANNs we chose a word embedding size of 256, an $L2$-regularization weight of $10^{-3}$, a dropout equal to 0.7, an MLP with a hidden layer of 32 neurons for the similarity score between clause and memory, and 64 neurons for the final classification layer. For strong supervision, we set the margin $\gamma = 0.5$ to ensure that uncorrelated legal explanations are irrelevant for classification, while we tuned the penalty coefficient separately for each unfairness category.

\begin{table}
    \centering
    \begin{tabular}{c|c|c|c|c|c|}
    \cline{2-6}
    & \multicolumn{5}{c|}{\textbf{Categories}} \\
    \cline{1-6}
    \multicolumn{1}{|c|}{\textbf{Model}} & A & CH & CR & LTD & TER \\
    \hline
    \multicolumn{1}{|c|}{SVM}  & 0.350 & \bftab{0.673} & 0.538 & 0.636 & 0.636 \\
    \multicolumn{1}{|c|}{CNN} & 0.361 & 0.654 & 0.584 & 0.627 & 0.612 \\
    \multicolumn{1}{|c|}{LSTM} & 0.326 & 0.639 & 0.498 & 0.589 & 0.589 \\
    \multicolumn{1}{|c|}{MANN (WS)} & 0.503 & 0.670 & 0.596 & 0.649 & 0.664 \\
    \multicolumn{1}{|c|}{MANN (SS)} & \bftab{0.526} & 0.665 & \bftab{0.606} & \bftab{0.659} & \bftab{0.666} \\
    \hline
    \end{tabular}
    \caption{Classification performance of the employed models according to macro-F1 computed on 10-fold cross-validation for unfair examples. For MANN, WS and SS stay for weak and strong supervision, respectively.\protect\footnotemark}
    \label{tab:f1_score}
\end{table}

\footnotetext{The reported performance is not directly comparable with the work of \cite{lagioia2019deep} for two different reasons: (1) the corpus is different, as it consists on a larger (and more heterogeneous) collection of 100 contracts; (2) the task here is a binary classification for each separate category, whereas \cite{lagioia2019deep} addressed the (simpler) problem of detecting potentially unfair clauses (of any category).}

\begin{table}
    \centering
    \begin{tabular}{|c|c|c|c|c|c|c|c|}
    \hline
    \textbf{Model} & \textbf{U} & \textbf{C} & \textbf{CP} & \textbf{CP@1} & \textbf{CP@3} & \textbf{MP} & \textbf{APM} \\
    \hline
    \multicolumn{8}{|c|}{Arbitration (A)} \\
    \hline
    Baseline@1    & 1.0 & 0.698 & 0.698 & 0.698 & 0.698 & 0.698 & 0.125 \\
    Baseline@2   & 1.0 & 0.358 & 0.358 & 0.358 & 0.358 & 0.358 & 0.125 \\
    MANN (WS)    & 0.292 & 0.283 & 0.968 & 0.548 & 0.839 & 0.968 & 0.839 \\
    MANN (SS)    & 0.547 & 0.500 & 0.914 & 0.741 & 0.897 & 0.862 & 0.500 \\
    \hline
    \multicolumn{8}{|c|}{Unilateral changes (CH)} \\
    \hline
    Baseline@1    & 1.0 & 0.837 & 0.837 & 0.837 & 0.837 & 0.837 & 0.143 \\
    Baseline@2     & 1.0 & 0.212 & 0.212 & 0.212 & 0.212 & 0.212 & 0.143 \\
    MANN (WS)           & 0.526 & 0.445 & 0.845 & 0.210 & 0.757 & 0.459 & 0.752 \\
    MANN (SS)           & 0.872 & 0.805 & 0.913 & 0.850 & 0.877 & 0.693 & 0.454 \\
    \hline
    \multicolumn{8}{|c|}{Content removal (CR)} \\
    \hline
    Baseline@1     & 1.0 & 0.546 & 0.546 & 0.546 & 0.546 & 0.546 & 0.059 \\
    Baseline@2    & 1.0 & 0.435 & 0.435 & 0.546 & 0.546 & 0.435 & 0.059 \\
    MANN (WS)           & 0.074 & 0.051 & 0.688 & 0.250 & 0.563 & 0.688 & 0.482 \\
    MANN (SS)          & 0.148 & 0.093 & 0.625 & 0.375 & 0.500 & 0.625 & 0.292 \\
    \hline
    \multicolumn{8}{|c|}{Limitation of liability (LTD)} \\
    \hline
    Baseline@1     & 1.0 & 0.676 & 0.676 & 0.676 & 0.676 & 0.676 & 0.056 \\
    Baseline@2    & 1.0 & 0.423 & 0.423 & 0.423 & 0.423 & 0.423 & 0.056 \\
    MANN (WS)          & 0.176 & 0.144 & 0.818 & 0.264 & 0.564 & 0.818 & 0.581 \\
    MANN (SS)          & 0.174 & 0.145 & 0.835 & 0.321 & 0.624 & 0.835 & 0.586 \\
    \hline
    
    \multicolumn{8}{|c|}{Unilateral termination (TER)} \\
    \hline
    Baseline@1     & 1.0 & 0.388 & 0.388 & 0.388 & 0.388 & 0.388 & 0.036 \\
    Baseline@2    & 1.0 & 0.229 & 0.229 & 0.229 & 0.229 & 0.229 & 0.036 \\
    MANN (WS)          & 0.0 & 0.0 & 0.0 & 0.0 & 0.0 & 0.0 & 0.0 \\
    MANN (SS)           & 0.998 & 0.631 & 0.632 & 0.387 & 0.492 & 0.499 & 0.136 \\
    \hline
    \end{tabular}
    \caption{Memory 
    statistics. Several metrics are reported in order to evaluate to what extent legal rationales are of use and to exclude ill-based performance. In particular, the following metrics are considered: memory usage (U), correct memory usage over unfair examples (C) and over examples for which memory is used (CP), along with a more fine-grained ranking version (CP@1-3), correct classification when memory is employed (MP) and, lastly, average memory usage per sample (APM). Models are compared with baselines that always select the most (@1) or second most (@2) frequent legal explanation.}
    \label{tab:memory_statistics}
\end{table}

\begin{scriptsize}
\begin{table}
\begin{tabular}{|>{\centering\arraybackslash}m{.05\linewidth}|>{\centering\arraybackslash}m{.54\linewidth}|>{\arraybackslash}m{.3\linewidth}|}
\hline
\textbf{Cat.} & \textbf{Clause} & \textbf{Targets/SS/WS}  \\ \hline
A & You and Airbnb mutually agree that any dispute,  claim or controversy arising out of or relating to these terms or the breach, termination, enforcement or interpolation thereof, or to the use of the Airbnb platform, the host services, or the collective content (collectively, "disputes") will be settled by binding arbitration (the "arbitration agreement"  &  \makecell[l]{\textbf{Targets}: \texttt{exceptions\_apply} \\ \textbf{SS}: \texttt{exceptions\_apply}\\ \textbf{WS}: |  } \\
\hline
CH & {[}...{]}, additionally, there may be times when we need to remove or change features or  functionality of the service or stop providing a service or access to third-party apps and services altogether & \makecell[l]{\textbf{Targets}: \texttt{anyreason}\\ \textbf{SS}:  \texttt{anyreason}\\ \textbf{WS}: \texttt{anyreason}, \texttt{nowarning},\\ \texttt{justponed}, \texttt{imbalance},\\ \texttt{consresp}, \texttt{againsterms},\\ \texttt{lawchange}, \texttt{update} } \\ \hline
\end{tabular}
\caption{Qualitative evaluation of strong supervision concerning two examples of different unfairness categories. The leftmost colum shows the category. The rightmost column reports the target set (\textbf{Targets}) associated to given clauses and the  legal rationales selected by the model, during the inference of two MANNs trained with strong (\textbf{SS}) and weak  supervision (\textbf{WS}). In these examples, the contribution of the strong supervision leads to more content-aware and selective classification, providing at the same time a qualitative tool for model interpretability.}
\label{tab:examples}
\end{table}
\end{scriptsize}

We present two different tables for results. In Table \ref{tab:f1_score} we report the standard classification metrics, namely the macro-average $F_1$ score computed on the cross-validation procedure, evaluated on the five unfairness categories of interest. 
Table \ref{tab:memory_statistics} instead reports several statistics about memory usage and related classification performance when considering both weak and strong supervision. In particular, we compute the following metrics: \begin{description} \item[\textbf{memory usage (U)}:] the percentage of input examples (clauses) for which memory is used \item[\textbf{coverage (C)}:] the percentage of correct memory slot selection over all unfair examples \item[\textbf{coverage precision (CP)}:] the percentage of correct memory slot selection over examples where memory is used. \end{description} These metrics allow to ascertain whether knowledge integration is effective, and to what extent. Moreover, we report a classification-oriented score concerning prediction accuracy with non-zero memory usage, namely memory precision (MP).
However, more detailed information about memory selection is required so as to exclude ill-behaved scenarios, such as where the model uses all available memory, and to avoid rushed positive conclusions about memory usage even though the model mostly prefers non-target justifications. To this end, we also report metric CP when only the memory slot with the highest score---that is, the most preferred explanation---is considered (CP@1) as well as the three slots with the highest scores are considered (CP@3). Finally, we report the average percentage of used memory, that is, the average number of memory elements selected with respect to total memory size (APM). 

The results in Table \ref{tab:f1_score} show that even a naive combination of a simple MANN architecture and raw knowledge representation is sufficient to show increased performance over traditional knowledge-agnostic models, such as the proposed neural baselines and the current state-of-the-art SVM solution. However, solely relying on classification performance is a non-exhaustive evaluation criterion to assess whether a model is better than another. Weak supervision is already sufficient to slightly enhance performance, whereas strong supervision still adds a margin of improvement. More specifically, harder unfair detection scenarios, such as A and CR, show a large benefit of added knowledge that compensates the high class distribution unbalance of these settings. However, a mixture of different factors, such as noisy knowledge representation, a high amount of possible unfairness explanations and the increase of available data, stand as an important challenge for knowledge integration, hence the performance improvement over the other categories of interest appears to be only moderate. In particular, LTD and TER categories present the highest amount of explanation variation, which increase the complexity of knowledge formulation for correct performance boost. 

As for the accuracy in providing explanations, which is the main point of adopting a MANN architecture, Table \ref{tab:memory_statistics} confirms our intuition that strong supervision provides a crucial element in correctly linking potentially unfair clauses to their correct rationales.
%
%
Table \ref{tab:memory_statistics} clearly shows that, when strong supervision is applied, the CP metric is very high, ranging from 0.63 (CR and TER) up to 0.8 and 0.9 (A, CH, LTD). In fact, the top-scoring rationale is very often correct for A and CH (above 0.7 and 0.8, respectively). For all the categories, the correct explanation is among the top-3 scoring in over 50\% of the cases (around 0.9 for A and CH). The APM metric also show that the MANN is quite selective in choosing the memory: in fact, the percentage of used explanations never exceeds 50\% of the memory size, on average.

Lastly, to further illustrate the benefits introduced by strong supervision, Table \ref{tab:examples} shows two examples of proper and selective memory usage that lead to correct unfairness detection. In particular, not only the model encouraged to use target legal rationales for given example, but it also learns to attentively use such added knowledge. 
On the other hand, a model without such direct supervisions may fail to filter out irrelevant information or to select relevant information for classification.

\section{Conclusions} \label{sec:conclusion}

A recent trend in legal informatics is the use of machine learning methods for document analysis and classification. Yet, especially in this domain, before the results of automated classifiers can be trusted, they need to be explainable. 

A pilot study on the use of MANN for detecting and explaining unfair clauses in consumer contracts was recently presented by  \cite{lagioia2019deep}, with promising results. We have extended that study by considering methods for better incorporating domain expertise into the explanations (rationales) produced by the MANN. Our results indicate that strong supervision could lead to more content-aware and selective classification of rationales. 

Our work also confirms that the problem of incorporating external knowledge into data-driven methodologies is a challenging one, particularly in the domain of legal analysis where context knowledge is significant, broad and noisy. An additional issue is scalability. Being able to rely on several sources of explanation may provide high flexibility, easing cross-domain adaption. However, incorporating large sets of legal rationales certainly requires approximation methods at memory addressing level \cite{chandar2016hierarchical, munkhdalai2019metalearned}, 
depending on the given model architecture. 

In the future we plan to consider other MANN architectures  \cite{xiong2016dynamic, miller2016key} and more advanced language models such as BERT  \cite{devlin2018bert}, and to investigate the applicability of our methodology to other areas of legal analytics, for example legal argument analysis  \cite{lippi2015claim} and privacy policies  \cite{contissa2018automated}. In particular, in the latter scenario, the high structural complexity demands richer task formulations rather than just sentence-level information: intra-document contextual information is crucial to assess the correctness of a legal statement and, thus, will represent one of the major leading research directions in the future.


\bibliographystyle{unsrt}      

\bibliography{ms}

\end{document}